\documentclass[
    ,final            
  ]
  {aipproc}

\layoutstyle{8x11double}

\begin{document}

\title{Quantum Spacetime, from a Practitioner's Point of View}


\author{J.\ Ambj\o rn}{
  address={The Niels Bohr Institute, Copenhagen University,
Blegdamsvej 17, DK-2100 Copenhagen \O , Denmark},
altaddress={Radboud University Nijmegen,
Institute for Mathematics, Astrophysics and Particle Physics,
Heyendaalseweg 135, 6525 AJ Nijmegen, The Netherlands}
}

\author{S.\ Jordan}{
  address={Radboud University Nijmegen,
Institute for Mathematics, Astrophysics and Particle Physics,
Heyendaalseweg 135, 6525 AJ Nijmegen, The Netherlands}
}

\author{J.\ Jurkiewicz}{
  address={Institute of Physics, Jagellonian University,
Reymonta 4, PL 30-059 Krakow, Poland}
}

\author{R.\ Loll}{
  address={Radboud University Nijmegen,
Institute for Mathematics, Astrophysics and Particle Physics,
Heyendaalseweg 135, 6525 AJ Nijmegen, The Netherlands}
}

\begin{abstract}
We argue that theories of quantum gravity constructed with the help of 
{\it (Causal) Dynamical Triangulations} have given 
us the most informative, quantitative models to date of {\it quantum spacetime}. Most importantly, 
these are derived {\it dynamically} from nonperturbative and background-independent quantum
theories of geometry. In the physically relevant case of four spacetime dimensions, the ansatz of
Causal Dynamical Triangulations
produces -- from a fairly minimal set of quantum field-theoretic inputs -- an emergent 
spacetime which macroscopically looks like a de Sitter universe, and
on Planckian scales possesses unexpected quantum properties. Important in deriving these
results are a regularized version of the theory, in which the quantum dynamics is well defined,
can be studied with the help of numerical Monte Carlo methods and extrapolated to
infinite lattice volumes.
\end{abstract}

\maketitle


\section{Introduction}

What unites cosmologists and researchers working on quantum gravity is their wish to understand
the nature and physical properties of spacetime, and how they are influenced by and in turn influence
the matter located in space and time. Although the work of cosmologists is usually associated with
the behaviour of the universe on the very largest scales ($10^{25}m$ and beyond), and that of quantum 
gravitators with ultra-short distances of the order of the Planck length, $\ell_{\rm Pl}\approx 10^{-35}m$, 
there are nevertheless phenomena which conceivably require input from both.
These fall under the umbrella of what one may call "quantum spacetime phenomenology" 
\cite{amelino}, the (so far putative) observable 
consequences for cosmology and astrophysics of a quantum microstructure of
spacetime, described by a (so far unknown) quantum theory of gravitation. The challenges
one faces in developing such a phenomenology have to do with the incredible 
weakness of the gravitational forces and the associated smallness of the characteristic Planck scale
at which quantum effects are expected to become important.
Conversely, in the absence of any obvious phenomenology we have little reliable guidance in our
search for a viable theory of quantum gravity. 

Well-known obstacles must be overcome
when trying to quantize classical general relativity, because of its highly nonlinear character and
the dimensionality of the gravitational coupling or "Newton's" constant, which implies that
a perturbative quantization of gravity becomes useless as one approaches Planckian distances.
At the same time, {\it non}perturbative candidate theories for quantum gravity, 
which provide more or less concrete 
quantum models of spacetime and the gravitational excitations at the Planck scale, face great 
difficulties in deriving any predictions -- observable or not -- because of their current 
incompleteness
and a scarcity of quantitative computational tools to bridge the gap
to more accessible macroscopic scales. This usually makes it difficult to verify that
one's favourite quantum-gravitational theory reproduces any aspects of the classical theory (as it must
in a suitable large-scale limit), let alone predict specific quantum corrections to Einstein's theory. 

Despite these daunting prospects, there is a lot of activity and exciting progress to report about in this area. The 
development and further sophistication of nonperturbative computational tools has in recent times enabled
us to derive quantitative results (yes, numbers!) in full\footnote{that is, without making any additional
symmetry assumptions to reduce the degrees of freedom drastically, as is done, for example, in quantum cosmology},
four-dimensional quantum gravity. 
In so far as they relate to the classical limit, these numbers can be used to
rule out (or rule in) specific models as promising candidates for quantum gravity. In so far as
they concern $\hbar$-effects, they cannot as yet be related to any true quantum spacetime
phenomenology, but can already be used to distinguish between different candidate
theories. In the ansatz of {\it Causal Dynamical Triangulations}, a
prominent example of such a theory, 
one meets concrete instances of both situations. 
The use of effective computational methods has also
exhibited a number of generic pitfalls in constructing a nonperturbative quantum gravity theory, which
could not have been anticipated from either the classical or the perturbative theory, thus giving us
valuable insights into what works and what does not.

\section{Quantum Spacetime}

\begin{figure}\label{foam}
  \includegraphics[height=.3\textheight]{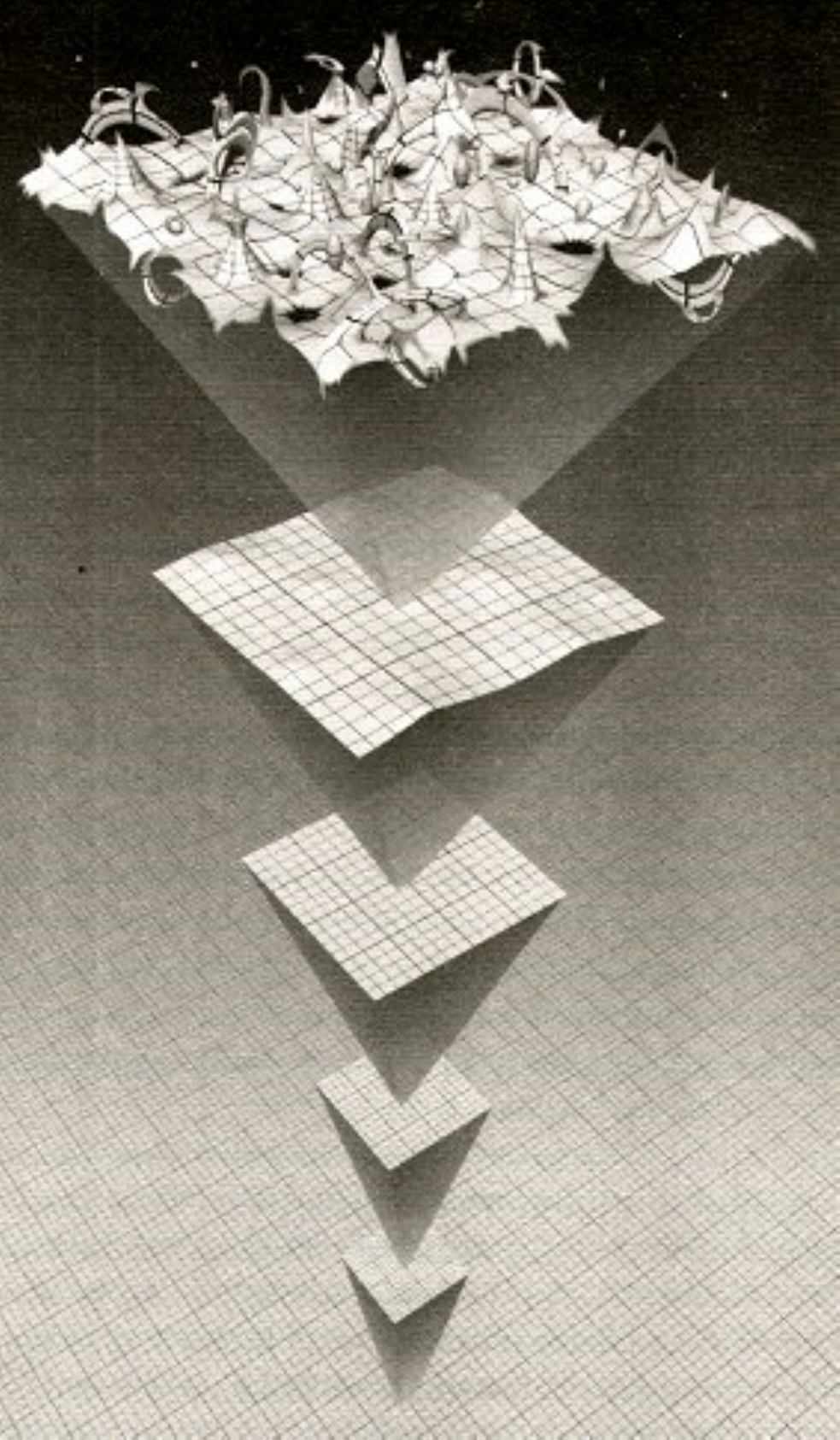}
  \caption{Zooming in on a piece of empty spacetime, which near the Planck scale one is 
  supposed to imagine as a frothing quantum foam.}
\end{figure}

Our working definition of a {\it quantum spacetime} is a "spacetime" (in the most general sense)
with quantum properties on Planckian scales, which in a suitable macroscopic limit can be
approximated by a classical curved spacetime $(M^{(4)},g_{\mu\nu}(x))$, consisting of a differentiable
manifold $M^{(4)}$ and a metric tensor $g_{\mu\nu}(x)$ of Lorentzian signature. We do not
know a priori which of the geometric or other properties of a classical spacetime will continue 
to be meaningful at the Planck scale; this is precisely one of the interesting questions the quantum
theory should be able to answer.\footnote{A good measure of how little we know about 
"quantum spacetime" is how few pictures there are on Google, where one finds just a handful of 
artistic impressions of "spacetime foam",  like that shown in Fig.\ \ref{foam}.}

There are essentially two ways to arrive at a quantum spacetime, a kinematical and
a dynamical one. In the first case, one chooses a particular classical 
("background") geometry 
and "embellishes" it with quantum fluctuations or, alternatively, arranges some microscopic
quantum degrees of freedom such that they "approximate" a given classical background
geometry. This can be thought of as a generalization of standard perturbative quantum
gravity, with the difference that the assumed Planckian quantum fluctuations will typically 
look very different from linearized quantum fields.
Since many researchers believe in a fundamental discreteness
of spacetime at short distances, these degrees of freedom will often be of a discrete
nature, involving sets of points or one-dimensional lattice-like structures, possibly with
additional relations or labels. Other examples of kinematical, "fuzzy" spacetimes are
given by certain noncommutative spaces.

The other, much harder way of obtaining a quantum spacetime is as the solution to the
dynamics of a full theory of quantum gravity, say, as the extremum of some path integral
or a solution to some Wheeler-DeWitt equation. The crucial difference with the kinematical
approach is that {\it no} distinguished background spacetime is put in by hand. Instead, similar to
what happens in the classical theory, a physical (quantum) spacetime is obtained only by
{\it solving} (quantum) equations of motion. 

Contrary to what the label "background-independent
quantization" may suggest, there are specific {\it choices} involved in
trying to set up such a nonperturbative quantum theory of gravity, most importantly
of (i) the microscopic degrees of 
freedom, representing whatever becomes of "gravity" at the Planck scale, 
(ii) the kinematical principles (causality, locality, etc.), symmetries and other algebraic structures, and
(iii) a dynamical principle -- whatever constitutes the "quantum equations of motion". 
There is little way of telling which are the correct choices and to what extent they will influence the
final outcome, without actually constructing the theories and computing some concrete numbers
from them. 

The trouble with a "kinematical" spacetime is that one cannot estimate to what extent
it approximates a true, dynamical quantum spacetime of the full theory before having solved the latter.
In other words, one does not know whether there is a nonperturbative theory of quantum gravity 
which has it as a stable solution or ground state. 
In the absence of dynamical solutions, kinematical quantum spacetimes may serve as useful 
models of the type of "quantum effects" 
that {\it may} be present, and one can try to study their phenomenological implications. 
Almost all quantum spacetimes that have been studied are kinematical and/or rely on an
a priori symmetry reduction in the dynamical degrees of freedom, where the symmetry is
suggested by that of a particular classical solution (for example, homogeneity, isotropy or spherical symmetry).
This implies that they carry a considerable degree of classical background structure {\it by
construction}, which makes it rather unsurprising if one can "rederive" a classical spacetime from
them in a suitable large-scale limit. 

This is very different from a truly background-independent derivation, where the "emergence" of 
classical spacetime -- if it can be shown to occur -- is highly nontrivial. As far as we are aware, 
the only fully nonperturbative
candidate theory of quantum gravity which has produced a quantum spacetime with genuine
classical large-scale properties is that of Causal Dynamical Triangulations (CDT), mentioned in the
Introduction, and to be discussed further below. 
Like any other formulation of quantum gravity, this approach involves working hypotheses on the
relevant ingredients (i)-(iii) above, which ultimately can only be justified {\it a posteriori}. 

Isn't this just as much guesswork as goes into the construction of a kinematical quantum spacetime?
The answer is {\it no}, because the presence of dynamics provides crucial additional information
about the system. Already the need for a quantum configuration space --
a domain of the path integral or a Hilbert space -- on which a dynamics is well defined 
(as opposed to being just "formal") imposes strong restrictions. Moreover, the
presence of dynamics exhibits which region(s) of phase space the quantum-geometrical
system is driven towards, and whether there are instabilities or other unexpected dynamical
features. It is not surprising that it was
in the context of {\it Dynamical} Triangulations (DT), which provides a concrete
calculational framework of this kind, that
generic, nonperturbative dynamical instabilities (associated with highly degenerate geometries)
of systems of 
strongly fluctuating higher-dimensional quantum geometry were first discovered 
and quantified \cite{degenerate1,degenerate2}.  

\begin{figure}\label{picasso}
  \includegraphics[height=.3\textheight]{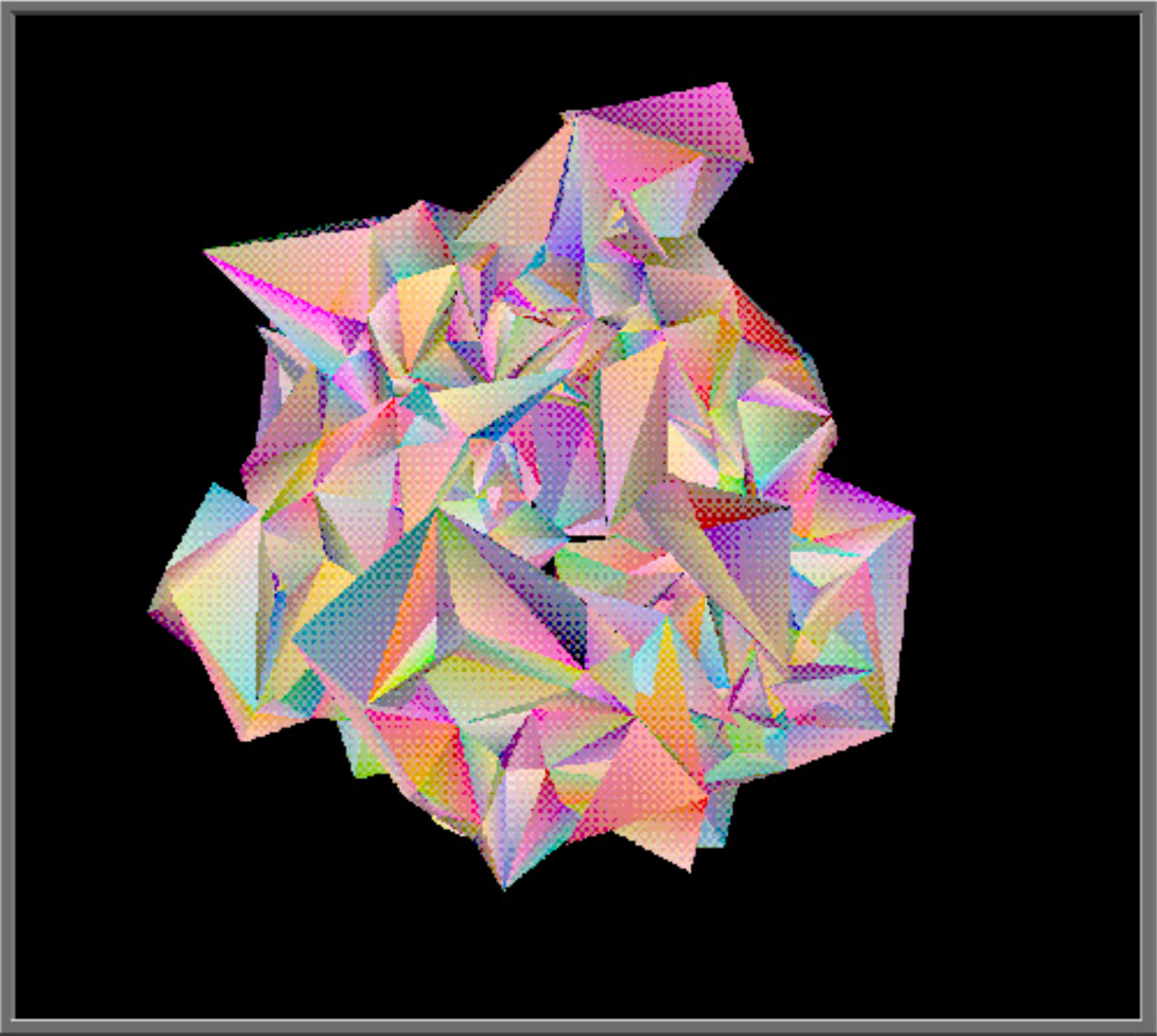}
  \caption{A "quantum spacetime", generated dynamically in the DT
  approach in two dimensions. Although it reflects certain
  arbitrary choices (here, that of a regularization in terms of flat triangles), it is a "typical" path 
  integral configuration and therefore 
 contains genuine {\it dynamical} information, for example, the fact that it in no way approximates
 a smooth spacetime locally. [source: P. Coddington]}
\end{figure}

A dynamical approach -- even before all of its implications are understood and analyzed -- can also give
valuable "pre-geometric" information about the quantum spacetime, for example, its 
dimensionality.\footnote{By depicting particular aspects of quantum geometry, both DT and CDT 
quantum gravity theories have produced pictures that are not only pretty, but convey
information about their {\it dynamical} content (Fig.\ \ref{picasso}).}
On Planckian scales, this may in principle deviate from the classical value of four. Again, one should be careful
to distinguish between genuine dynamical properties of the spacetime and those which are
simply consequences of some kinematical choices put in by hand. For example, if one
postulates that spacetime is fundamentally discrete and chooses a particular type of 
fundamental "building block" or quantum excitation whose size is the Planck length, 
all properties of the quantum spacetime {\it at that scale} will strongly depend on
this choice. This is not really good enough if one is interested in a fundamental description
of quantum gravity at the Planck scale, because in that case one still needs to argue why a particular
choice of discrete structure is better than the infinitely many other choices one could have made instead.

One way to obviate the need for seemingly arbitrary guesses about "physics at the Planck scale"
is to work with a regularization -- which {\it is} necessary at an intermediate stage to make 
calculations well defined -- whose associated cut-off scale is sent to zero in the end or, from a
practical point of view,
taken to be at least {\it significantly smaller} than 
the physical scale one is interested in describing. A well-known mechanism for removing 
the dependence on arbitrary regularization details is that of "universality", realized under favourable
conditions when approaching a phase transition point (of order 2 or higher) 
in the space of bare coupling constants
of a regularized statistical model underlying the candidate theory of quantum gravity under 
consideration. If an ultraviolet fixed point and an associated scaling limit\footnote{usually
defined in terms of a divergent correlation length of the theory} exist, the theory is said to possess a "continuum
limit". Despite the name, such a construction is as a matter of principle
not necessarily incompatible with 
fundamental discreteness. For example, it may happen that a distinguished length scale is
generated dynamically, and appears as the minimal eigenvalue of a quantum operator measuring
length, area or some other geometric quantity, similar to what happens in the kinematical 
sector of loop quantum gravity, say. However, there are to our knowledge no compelling arguments 
for the existence of a minimal length scale in quantum gravity, as opposed to 
wishfully thinking that it {\it should} exist to provide a "natural cut-off" to momentum 
integrals.\footnote{How to give an operationally well-defined meaning to the notion of a {\it minimal length}
in a nonperturbative theory of quantum gravity is of course yet another nontrivial issue.}

There is strong evidence that the scenario sketched above -- a regularized quantum field-theoretic
framework for gravity with a set of minimalist ingredients (i), (ii) and (iii), from which a full theory 
of quantum gravity emerges, along with a specific "quantum spacetime" as its ground state -- is indeed
realized in the ansatz of four-dimensional CDT. -- In the remainder of this
article, we will only give the briefest of summaries of the ingredients and some recent results
of this approach to quantum gravity. It should be read in conjunction with our companion article
\cite{szczecin2} in the same volume, which contains some further technical details. In what follows, we
will specifically highlight some new results concerning the
phase structure of CDT quantum gravity. For a much more
comprehensive exposition of this material and a more complete bibliography, we refer the interested reader to
several recent reviews of the subject \cite{cdtreview1,cdtreview2,cdtreview3}.  

\section{Quantum Spacetime from Causal Dynamical Triangulations}

\subsection{Tools and ingredients}

\begin{figure}\label{3dtriang}
  \includegraphics[height=.2\textheight]{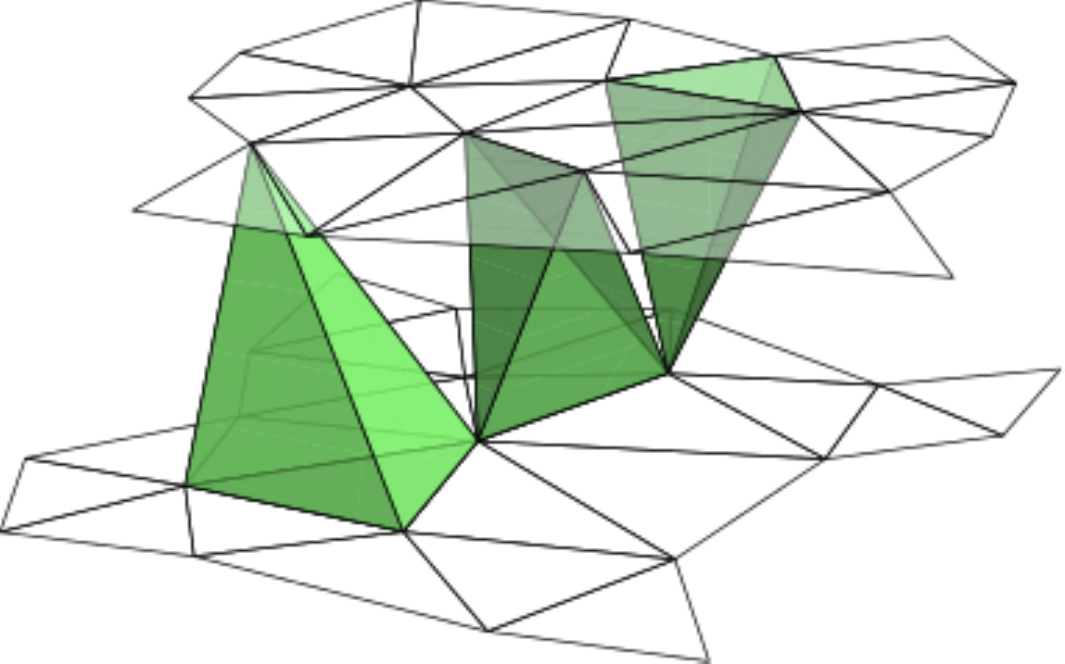}
  \caption{The building blocks of the Lorentzian triangulations are arranged in layers, in such
  a way that all vertices of the triangulation lie in spatial slices of constant integer
  proper time. For ease of visualization, the figure illustrates the analogous situation in 2+1 dimensions. 
  Shown are three simplicial building blocks, and the different ways in which their vertices can
  be distributed over two adjacent spatial slices, with proper time labels $t$ and $t+1$.
   [source: T. Budd]}
\end{figure}

"Quantum gravity from CDT" relies on few ingredients and a minimal set of prior assumptions.
In terms of the ingredients listed in the previous section, we have
\begin{enumerate}
\item[(i)] {\it microscopic degrees of freedom}: curved Lorentzian geometries of fixed topology, encoded in 
four-dimensional, piecewise flat triangulations ("simplicial manifolds") with curvature concentrated at 
two-dimensional sub-simplices (the triangles of the triangulation).
\item[(ii)] {\it kinematical principles and other structures}: quantum superposition principle, locality, causality; 
notions of proper time and Wick rotation.
\item[(iii)] {\it dynamical principle}: path integral over the domain (i) with weights given by the
Einstein-Hilbert action with a cosmological constant term, implemented nonperturbatively 
using standard tools from quantum field theory, including appropriate renormalization of couplings
in the limit as the ultraviolet lattice cut-off (edge lengths of the triangulations) is taken to zero.
\end{enumerate}
This minimalist and rather conservative starting point, following closely the dynamical ingredients of the classical
theory, without postulating any new symmetries or degrees of freedom, is all that is necessary to produce   
an array of truly interesting results. Key to obtaining an emergent classical geometry are
causality conditions imposed on the individual triangulations, with the help of a global proper-time
slicing \cite{firstpaper,causal}, such that the simplicial building blocks are arranged in layers, see Fig.\ \ref{3dtriang}.
This particular definition of the domain of the path integral is not "pulled out of a hat", but 
motivated by previous investigations, most importantly, in the much-studied Euclidean DT version
of the formulation.
The calculational set-up can be seen as the gravitational analogue of a lattice quantum field theory
(like lattice QCD), although not on a given fixed (hypercubic, say) lattice, but incorporating a sum 
over all inequivalent triangular lattice geometries, reflecting the dynamical nature of spacetime.
In addition, one can of course add other matter degrees of freedom by assigning field variables
to the elements of the simplicial lattices and including them in the path integral. Of crucial importance
in all cases is the availability of numerical methods to evaluate the regularized, nonperturbative 
path integrals (after Wick rotation), using Monte Carlo techniques and finite-size scaling. 
A wealth of results from studying DT and CDT models in dimensions two, three and four, with and 
without matter, have revealed a large degree of universality, the highly desirable property of
the continuum limits of these theories to be largely independent of the details of the implementation
and regularization. Universality covers various regularity conditions on the path integral configurations,
and, to a large degree, the form of the action and the measure, but {\it not} the signature of the geometries. 
Choosing Euclidean or Lorentzian signature leads to inequivalent continuum theories, as was
first demonstrated in two dimensional gravity \cite{firstpaper}, an early key insight obtained
by using dynamical triangulations.

\subsection{Some Results}

As a consequence of the regularization chosen and a subsequent Wick rotation, the original 
Lorentzian gravitational path integral is transformed into a statistical partition function of a specific
ensemble of Euclidean simplicial manifolds, with weights given by the Euclideanized Einstein-Hilbert
action, now expressed as a function of the variables characterizing a given triangulation, 
see \cite{szczecin2}, eqn.\ (3). The action, and therefore the partition function, depends on
three bare coupling constants, the cosmological constant $\kappa_4$, the inverse Newton constant $\kappa_0$
and a so-called asymmetry parameter $\Delta$, which simply is the proportionality factor
between the lattice link length in the time direction and that in the spatial directions ($\Delta=0$
corresponds to isotropic length assignments). The latter appears
naturally as a free parameter of the regularized theory, since the Lorentzian geometries are 
by construction {\it anisotropic} in space and time. Moreover, the associated proper-time slicing
and suppression of the generation of so-called baby universes (leading to spatial topology
changes and causality violation) appear to be necessary to obtain an emergent de Sitter space
on large scales \cite{desitter1,desitter2}. 

\begin{figure}\label{phasediagram}
  \includegraphics[height=.23\textheight]{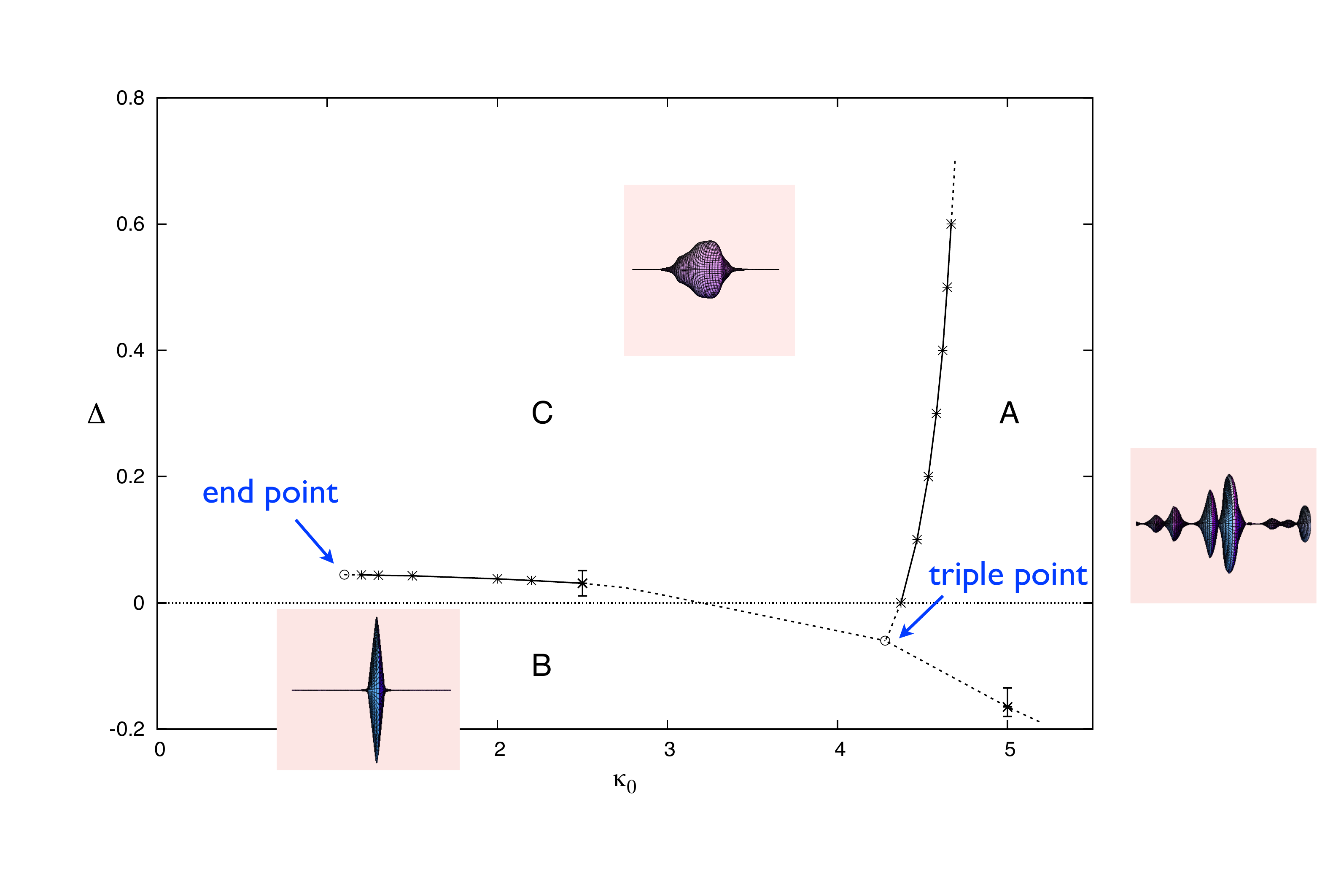}
  \caption{Phase diagram of four-dimensional CDT quantum gravity, as function
  of the inverse Newton constant $\kappa_0$ and the asymmetry parameter $\Delta$, measured at
  a spacetime volume of 84.000 four-simplices. The
  phases A, B and C can be distinguished by their typical volume profiles, 
  as indicated in the figure. The physical phase C contains extended de Sitter universes. 
  The A-C and B-C phase transitions have been found to be of first and second order 
  respectively \cite{phases1,phases2}.}
\end{figure}

It turns out that in all DT models, fine-tuning the bare cosmological constant to its critical value is tantamount
to taking the infinite-volume limit (in lattice units), a necessary step for obtaining a theory in the continuum.
This leaves us with two free parameters for the case of CDT quantum gravity in four dimensions, namely,
$\kappa_0$ and $\Delta$.\footnote{To address a frequent misconception, let us point out that our choice of 
the Einstein-Hilbert does not mean that
higher-curvature terms are ignored or suppressed. These will all be present in the path integral
and are contained in the "entropy" of path-integral configurations at a given value of the bare action. 
Not including higher-order curvature terms in the bare action only means that these terms will {\it not}
be associated with a tunable bare coupling constant in the full, effective action of the theory. That
this minimal assumption leads to an interesting continuum theory, by fine-tuning at most
$\kappa_0$ and/or $\Delta$, obviously needs to be verified.} The qualitative structure of the relevant
$\kappa_0$-$\Delta$-phase diagram (Fig.\ \ref{phasediagram}) has been known for 
some time \cite{phaseold}. 
There are three phases, A, B, and C, which can be distinguished with the help of an "order parameter",
given by the overall shape of the quantum spacetime that emerges as ground state in the given
phase. This so-called {\it volume profile} $V_3(t)$ is simply the three-volume of space (by definition
compact, with the topology of a three-sphere) at proper time $t$. The quantity $V_3(t)$ for a
given path integral geometry is of course not an observable, only its expectation value $\langle V_3(t)
\rangle$ in the ensemble is. Nevertheless, "typical" path integral histories in the different phases
exhibit very distinct characteristics, as illustrated by the three small figures included in the phase
diagram of Fig.\ \ref{phasediagram}. 
Only in phase C can the average volume profile $\langle V_3(t)\rangle$ be matched with 
that of a cosmological solution to the classical Einstein equations, namely, the de Sitter universe 
\cite{desitter1,desitter2}. This can be done with quite spectacular precision (c.f. Figs.\ 3 and 4 in \cite{szczecin2}),
including the quantum fluctuations, which can be matched to those computed from a related
minisuperspace model. 

In order to establish whether we can understand quantum gravity within the standard framework of
critical statistical-mechanical systems, where continuum limits are associated with critical points at
second- or higher-order transitions, we have recently performed a detailed analysis of the phase
transition lines bordering the physical phase C. The numerical determination of the order of the phase
transitions is a subtle affair, because of the characteristic slow-down of the simulations near the
transition, which make the simulations computationally very expensive. One also has to determine
which standard indicators of the order of a phase transition can be adapted to the case of fluctuating geometry,
and how. Only by combining several such criteria have we been able to determine the order of the
phase transitions with confidence \cite{phases2}. 

\begin{figure}\label{histogram}
  \includegraphics[height=.25\textheight]{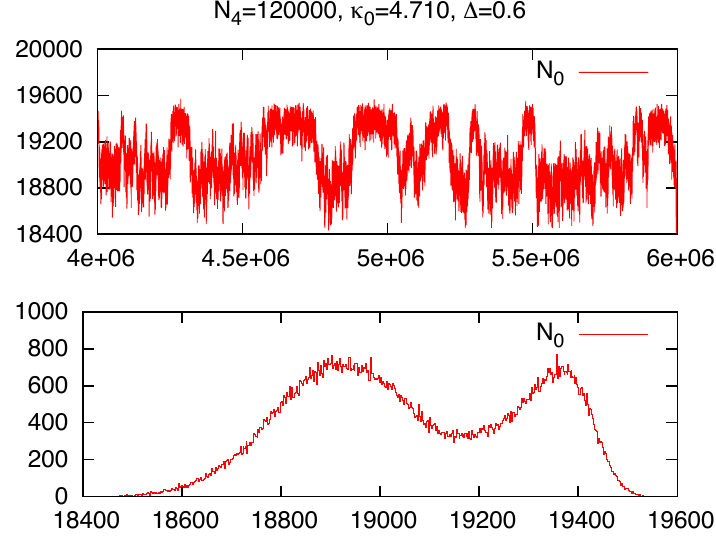}
  \caption{Monte Carlo time evolution of the quantity $N_0$ (number of vertices of the
  triangulated spacetime) conjugate to the
  coupling constant $\kappa_0$ in the bare CDT action at the A-C transition (top),
  together with the associated histogram, with characteristic double-peak structure (bottom).}
\end{figure}

More specifically, we have (i) used a histogram analysis
of selected quantities conjugate to the couplings controlling the approach to the respective transition lines, 
(ii) extracted so-called shift exponents from measuring the location of the maxima of the
susceptibilities associated with these quantities, and (iii) measured a variety of Binder 
cumulants. The histograms exhibit a clear two-peak structure near either transition line of the 
CDT system (see Fig.\ \ref{histogram}), 
the pertinent question being whether in the limit of large volumes they fuse into
a single one (indicative of a second-order transition) or become more pronounced (in line with
a first-order transition). 

The combined outcomes of these measurements constitute strong evidence that the A-C transition, between
the oscillatory and de Sitter phase, is first order, and the B-C transition between the "time-collapsed"
and the physical de Sitter phase is second order. This finding is potentially very significant, because
all points along a second-order line are potential candidates for the existence of continuum
theories. We also note two distinguished points along the B-C line, namely, an apparent end point for small
$\kappa_0$, as well as a triple point where all three phases meet. The finding of a second-order phase
transition, so far unique in four-dimensional quantum gravity, suggests that in order to probe
quantum spacetime at even smaller scales than what has been achieved until now -- where the de Sitter universes
have had a linear extension of about 12-20 Planck lengths across -- one needs to perform simulations
closer to the B-C transition line. Making the algorithms more efficient in the neighbourhood of this transition line
presents a considerable technical challenge and is currently under investigation. Physical statements extracted
for the infinite-volume limit based on measurements taken inside phase C, especially those referring to 
large-scale structure, are not invalidated by this result. However, we have noted before that our current 
"resolution" is not sufficient to analyze certain true Planck-scale details of geometry like, for example, deviations
of the volume profile $\langle V_3(t)\rangle$ near its end points, where $V_3(t)$ is small and quantum 
corrections to the potential energy of the relevant minisuperspace model should come into play.

\section{Summary and Outlook}
After discussing the status of "quantum spacetime", and possible ways of obtaining it in a theory
of quantum gravity, we have focussed on the approach of Causal Dynamical
Triangulations, which arguably has made the most progress in demonstrating that and how
a solution to the classical Einstein equations can emerge from a nonperturbative and 
background-independent formulation in a large-scale limit. The investigation of this candidate
theory of quantum gravity is still ongoing. The results obtained so far on the {\it short-scale},
Planckian structure of the theory are quite remarkable, because they have revealed 
totally unexpected (from a perturbative point of view) properties, most prominently that of
dimensional reduction, as described in the companion paper \cite{szczecin2} (see especially
Fig.\ 5 there, illustrating the scale-dependent spectral dimension of quantum spacetime). 
CDT, together with its nonperturbative computational
toolbox, gives us unprecedented glimpses of this highly nonclassical regime, which so far
cannot be accessed reliably by other methods. Other insights include the pivotal role played
by "entropy", the number of microscopic realizations for a given value of the action, which --
amongst other things -- is responsible for curing the problem of unboundedness of the
path integral and the associated instability \cite{cdtreview3}. 

To demonstrate that the quantum theory obtained
from CDT is the correct (and hopefully unique) theory of quantum gravity, further aspects
of the classical theory need to be "rederived", and quantum properties at Planckian scales
to be probed in more detail. All of these require the identification and spectral analysis 
of suitable "observables", on top of the already known ones. Finding observables which
are well defined in absence of any background geometry, as is the case in a nonperturbative
formulation, is of course one of {\it the} central challenges of any approach to quantum gravity. The
advantage of the CDT formulation discussed here is that the issue takes on a concrete form,
thanks to the presence of a well-defined computational set-up where observables can be operationally
implemented, measured and interpreted. In quantum gravity, this is just about as good as it comes.

\begin{theacknowledgments}
JJ and RL are most grateful to the organizers of Multicosmofun '12, and to M.P Dabrowski in particular,
for their extraordinary hospitality in Szczecin. JA thanks the Danish Research Council for financial
support via the grant "Quantum gravity and the role of black holes", and the EU for support through the ERC Advanced
Grant 291092, "Exploring the Quantum Universe" (EQU). SJ and RL acknowledge support through a
Projectruimte grant by the Dutch Foundation for Fundamental Research on Matter (FOM).
JJ acknowledges partial support of the
International PhD Projects Programme of the Foundation for Polish Science within the European Regional
Development Fund of the European Union, agreement no. MPD/2009/6.
\end{theacknowledgments}







\end{document}